\documentclass[twoside,english]{elsarticle}
\usepackage[T1]{fontenc}
\usepackage[latin9]{inputenc}
\usepackage{graphicx}

\makeatletter

\usepackage{amsfonts}
\usepackage{amsmath}

\makeatother

\usepackage{babel}
\begin{document}

\begin{frontmatter}{}

\title{Typical event horizons in AdS/CFT}

\author{Steven G. Avery}

\author{David A. Lowe}

\address{Department of Physics, Brown University, Providence, RI 02912, USA}
\begin{abstract}
We consider the construction of local bulk operators in a black hole
background dual to a pure state in conformal field theory. The properties
of these operators in a microcanonical ensemble are studied. It has
been argued in the literature that typical states in such an ensemble
contain firewalls, or otherwise singular horizons. We argue this conclusion
can be avoided with a proper definition of the interior operators.
\end{abstract}

\end{frontmatter}{}

\section{Introduction}

For black holes in asymptotically anti-de Sitter spacetime there are
two natural choices of vacua compatible with the symmetries. One such
vacuum is the analog of the Boulware vacuum \citep{Boulware:1974dm}:
positive frequency field modes far from the black hole, defined with
respect to the timelike Killing vector, annihilate the vacuum state.
Another natural choice is the analog of the Hartle--Hawking vacuum
\citep{Hartle:1976tp}, where positive frequency is defined with respect
to time translations of smooth global slices. This choice of vacuum
gives rise to entanglement between the left and right asymptotic regions
of the maximally extended AdS--Schwarzschild Penrose diagram and balanced
thermal fluxes of ingoing and outgoing modes.

The canonical and microcanonical ensembles of such black holes were
first studied by Hawking and Page \citep{Hawking:1982dh}. They reached
the important conclusion that for sufficiently large black holes,
relative to the AdS radius of curvature, the ensembles are thermodynamically
stable. In this limit, the horizon entropy of the black hole dominates
versus thermal excitations of matter fields. It is this limit that
is of interest in the present work.

More recently, Marolf--Polchinski (MP) \citep{Marolf:2013dba} studied,
using the microcanonical ensemble, the number operator for Kruskal-like
modes, those natural from the viewpoint of a freely falling observer.
They argued that in the holographic approach to quantum gravity, in
a typical eigenstate of such normalizable modes, this number operator
would always be of order 1. This then implies that typical black holes
always have violations of general covariance near the horizon. In
the following we re-examine this analysis and find that with an alternate
construction of the interior observables, this conclusion can be avoided.

\section{Semi-classical approach\label{sec:Semi-classical-approach}}

\begin{figure}
\includegraphics[scale=0.3]{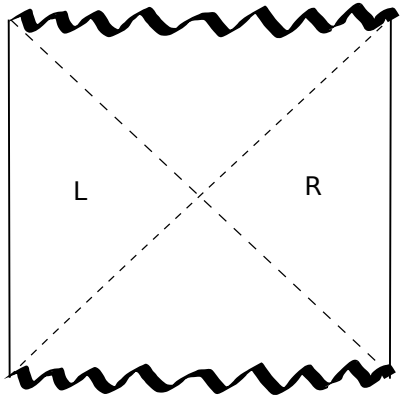}\protect\caption{\label{fig:The-Penrose-diagram}The Penrose diagram for Schwarzschild-anti
de Sitter spacetime.}
\end{figure}
Let us review the general setup. Following MP, we wish to consider
the microcanonical ensemble of a CFT with a holographic asymptotically
AdS description. The ensemble is defined by fixed mass $M$. We choose
$M$ sufficiently large that black holes give the dominant contribution
to the entropy in the bulk. Thus, we expect the bulk description (up
to exponentially suppressed corrections) to be a mass $M$ asymptotically
AdS black hole. Let us consider the effective field theory description
of the bulk. As is well-known, in curved backgrounds there is no preferred
mode decomposition of the field.

The $b$-modes are eigenstates of the timelike Killing vector at infinity,
which is closely associated with the CFT Hamiltonian. The Penrose
diagram for the anti-de Sitter Schwarzschild black hole is shown in
figure \ref{fig:The-Penrose-diagram}. While the setup we are considering
has a single CFT and a single asymptotic AdS spacetime, it is convenient
to introduce the full Schwarzschild spacetime and use the left patch
as a way to parametrize interior modes.

A bulk scalar field in the right patch (R) may be decomposed as
\[
\phi=\sum_{k}b_{R,k}\phi_{R,k}+b_{R,k}^{\dagger}\phi_{R,k}^{*}
\]
where $k$ schematically represents the set of labels. We refer to
these modes as the right $b$-modes, and they annihilate the Schwarzschild
vacuum state $b_{R,k}|0\rangle_{S}=0$. 

If the CFT description is to include a description of the black hole
interior, one must also consider a set of operators representing the
fields in the left patch (L), which propagate into the upper patch
in figure \ref{fig:The-Penrose-diagram}. Both sets of modes are needed
to provide a complete description of the field in the upper patch.
The full decomposition, valid in all coordinate patches is then
\[
\phi=\sum_{k}b_{L,k}\phi_{L,k}+b_{L}^{\dagger}\phi_{L,k}^{*}+b_{R,k}\phi_{R,k}+b_{R,k}^{\dagger}\phi_{R,k}^{*}
\]
and the Schwarzschild vacuum state is also annihilated by the left
$b$-modes $b_{L,k}|0\rangle_{S}=0$. Note that we define the mode
functions, $\phi_{L/R}$, in the above such that they only have support
in the appropriate (left/right) region.

One may also choose to decompose the field with respect to Kruskal
modes, which are analytic across the horizon
\[
\phi=\sum_{k}a_{k}\phi_{K,k}+a_{k}^{\dagger}\phi_{K,k}^{*}\,.
\]
We refer to these modes as the $a$-modes. These modes annihilate
the Kruskal/Hartle-Hawking vacuum $a_{k}|0\rangle_{K}=0$. These modes
look rather complicated when decomposed into frequencies with respect
to the timelike Killing vector at infinity. However as shown in \citep{Unruh:1976db},
these may be rewritten in terms of another set of operators $d_{L,k}$
and $d_{R,k}$ that annihilate $|0\rangle_{K}$ but are simply related
to the $b$-modes
\begin{eqnarray*}
\phi & = & \sum_{k}\left(2\sinh\left(\beta\omega_{k}/2\right)\right)^{-1/2}\left(d_{R,k}\left(e^{\beta\omega_{k}/2}\phi_{R,k}+e^{-\beta\omega_{k}/2}\phi_{L,-k}^{*}\right)\right.\\
 &  & +\left.d_{L,k}\left(e^{-\beta\omega_{k}/2}\phi_{R,-k}^{*}+e^{\beta\omega_{k}/2}\phi_{L,k}\right)\right)+h.c.
\end{eqnarray*}
where $\beta$ is the inverse Hawking temperature of the black hole
and $\omega_{k}$ is the positive frequency associated with the mode
labeled by $k$. These operators are related to the $b$-mode operators
by a Bogoliubov transformation 
\begin{eqnarray*}
b_{L,k} & = & \left(2\sinh\left(\beta\omega_{k}/2\right)\right)^{-1/2}\left(e^{\beta\omega_{k}/2}d_{L,k}+e^{-\beta\omega_{k}/2}d_{R,-k}^{\dagger}\right)\\
b_{R,k} & = & \left(2\sinh\left(\beta\omega_{k}/2\right)\right)^{-1/2}\left(e^{\beta\omega_{k}/2}d_{R,k}+e^{-\beta\omega_{k}/2}d_{L,-k}^{\dagger}\right)
\end{eqnarray*}
which allows the different vacua to be related via
\begin{eqnarray}
|0\rangle_{K} & = & \prod_{k}\exp\left(e^{-\beta\omega_{k}/2}b_{L,k}^{\dagger}b_{R,k}^{\dagger}\right)|0\rangle_{S}\nonumber \\
 & = & \prod_{k}\sum_{n_{k}=0}^{\infty}e^{-\beta\omega_{k}n_{k}/2}|N_{b,L,k}=n_{k}\rangle\times|N_{b,R,k}=n_{k}\rangle\label{eq:hartlehawking}
\end{eqnarray}
where $N_{b,L,k}$ and $N_{b,R,k}$ are the number operators for the
$b$-modes. 

The number operator relevant for an infalling observer can be defined
as the number operator built from the Kruskal mode number operators
\begin{equation}
N_{d,k}=d_{L,k}^{\dagger}d_{L,k}+d_{R,k}^{\dagger}d_{R,k}\label{eq:infallnumber}
\end{equation}
and this annihilates $|0\rangle_{K}$. In fact, everything we say
applies to each term in the above separately. We will utilize this
expression momentarily.

The MP argument instructs us to compute the ensemble average of the
expectation value of $N_{d,k}$. Since the $b$-modes' frequencies
are related to the CFT Hamiltonian, it is natural to evaluate the
ensemble average in $N_{b}$ eigenstates. It follows from the Bogolyubov
transformation that each $N_{b}$ eigenstate gives an $O(1)$ contribution
to the expectation value of $N_{d,k}$. Since the number operator
is positive definite there can be no cancellation in computing the
ensemble average. Since this applies to each Kruskal-like mode, one
concludes that the typical microstate of the ensemble has a firewall.
We return to this argument later.

\section{\label{sec:Euclidean-quantum-gravity}Euclidean quantum gravity approach}

The Euclidean Gravity framework imposes periodicity in imaginary time
to formulate the canonical ensemble \citep{Hawking:1982dh}. The microcanonical
ensemble is then defined via an inverse Laplace transform of the canonical
ensemble. In the gravitational sector, the correct Bekenstein--Hawking
black hole entropy is obtained. However in this approach the horizon
entropy arises from geometric factors, rather than from state counting.

In addition, there is a contribution due to a thermal bulk field modes.
This contribution can be viewed as computing the entropy of the reduced
density matrix obtained by starting with the pure state \eqref{eq:hartlehawking}
and tracing over the left-modes. For sufficiently large total energies,
the microcanonical ensemble is dominated by exclusively the black
hole entropy contribution, with a negligible term coming from thermal
field configurations outside the black hole horizon. This observation
will be important later, as it is necessary for a typical bulk mode
to be in the global vacuum state (i.e. Hartle--Hawking vacuum) in
order that a firewall not be seen.

\section{AdS/CFT Approach}

There is strong evidence the CFT is able to correctly reproduce the
Bekenstein--Hawking contribution to the entropy in the large mass
limit \citep{Gubser:1996de}. The entropy is reproduced by direct
state counting, up to an overall constant that is difficult to determine
precisely, because the CFT is strongly coupled in the limit that it
is dual to a gravitational phase.

This approach must also yield significant corrections to the approach
of section \ref{sec:Semi-classical-approach}. Let us focus on the
case of the four-dimensional bulk spacetime theory for the sake of
definiteness. The boundary of the theory is $S^{2}\times\mathbb{R}$,
with the $\mathbb{R}$ factor corresponding to the time coordinate.
Because the spatial sections are compact spheres, the energy spectrum
of the conformal field theory becomes discrete. This induces a particular
cutoff on the spectrum of the bulk theory.

\subsection*{Reconstructing bulk fields from CFT data}

The reconstruction of bulk fields in the eternal black hole geometry
was considered in \citep{Lowe:2009mq,Hamilton:2006fh,Hamilton:2007wj}.
There it is possible to build bulk fields in the interior of the horizon,
but at the price of realizing the bulk fields as operators in a ``doubled''
CFT, corresponding to the two asymptotic infinities of the eternal
black hole. To describe the microcanonical ensemble, one would like
to solve the problem for black holes with a single asymptotic exterior
region,corresponding to states in a ``single'' CFT. In this case
the methods of \citep{Lowe:2009mq} readily generalize only to the
exterior of the black hole. Let us consider various possible hypotheses
for realizing bulk operators in the black hole interior.

\subsubsection{\label{sub:No-interior-bulk}No interior bulk operators}

This is certainly a logical possibility in view of the semiclassical
spacetime of figure \ref{fig:The-Penrose-diagram}. The exterior region
exists for infinite time, and any CFT state should be mapped uniquely
into some configuration of the exterior bulk fields (and perhaps more
exotic configurations like strings). In particular, it should be a
good approximation to map CFT operators into operators that produce
some superposition of $b$-mode eigenstates. Since the exact spectrum
is discrete, one can try to map any such CFT state into some superposition
of only a finite number of $b$-mode eigenstates. This step is crucial
for the diagonalization argument in MP. As pointed out in \citep{Avery:2013vwa},
there is no reason the discretization induced by the CFT will map
to a finite number of $b$-mode eigenstates. In fact, the arguments
of the next subsection suggest they at least start out being well-represented
by superpositions involving infinite numbers of $b$-mode eigenstates
as in \eqref{eq:hartlehawking}. In essence, we claim that diagonalizing
$N_{b}$ and going to the microcanonical ensemble may not be compatible,
in view of the need to cut off the effective field theory description
in some way. Some may view this as a breakdown of the bulk effective
field theory, and thus consistent with the MP argument; however, following
Lowe and Thorlacius \citep{Lowe:2014vfa} we claim that the breakdown
need not have severe consequences for reasonable observables.

\begin{figure}
\includegraphics[scale=0.2]{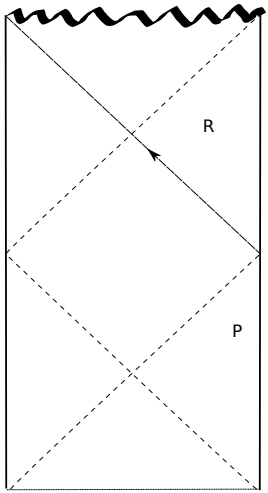}\protect\caption{\label{fig:A-black-hole}A black hole with a single exterior region.
The upper dashed line shows the global horizon. The infalling matter
is denoted by the line with the arrow.}
\end{figure}

\subsubsection{\label{sub:Entangled-bulk-fields}Entangled bulk fields}

The previous picture fails if one takes into account the recurrence
of bulk field configurations due to the discreteness of the exact
spectrum. For a quantum state, one expects a Poincare recurrence (i.e.
for some set of observables to exhibit a quasi-periodicity) on some
timescale ranging from the Heisenberg time $e^{S}$ (at a minimum)
to of order $e^{ce^{S}}$ for some constant $c>0$. If one combines
this with an ergodic hypothesis, then over such large timescales,
one expects to observe time intervals when the horizon of the black
hole disappears and then reforms due to collapse. 

In this situation, as shown in figure \ref{fig:A-black-hole}, we
can freely evolve fields in the region inside the global horizon back
to the past region P. In this region, one can apply the construction
of \citep{Lowe:2009mq,Hamilton:2006fh,Hamilton:2007wj} to reproduce
the bulk fields in terms of CFT operators. Thus we are free to time
evolve forward, to obtain local bulk fields inside the horizon.

At first sight, it seems this should be sufficient to time evolve
the bulk fields throughout the interior of the horizon. However this
is limited by the following consideration. One only expects the exact
theory to be able to reproduce local bulk fields on length scales
larger than the Planck length (at best). However to time evolve bulk
fields throughout the horizon interior, it is necessary to construct
bulk fields on much shorter length scales in region P. The horizon
red-shift factor then expands these scales, when viewed from the perspective
of late-time freely falling observers in the black hole interior.
This limits the time after formation of the black hole at which this
procedure will work to Schwarzschild times of order $\beta\log S,$
where $\beta$ is the inverse Hawking temperature and $S$ is the
Bekenstein-Hawking entropy. This timescale matches the scrambling
time obtained using different methods in \citep{Shenker:2013pqa}.

For pre-scrambling times after black hole formation, the interior
bulk fields are correlated with the exterior bulk fields, and one
expects to reproduce a good approximation to the results in the global
vacuum, modulo excitations set up near the horizon prior to formation.
From the CFT viewpoint, this means that the bulk operators cannot
produce states that are superpositions of finite numbers of $b$-mode
eigenstates---rather any state with these properties must be viewed
as some superposition of AdS global mode eigenstates, each of which
can only be represented as a superposition of an infinite number of
$b$-mode eigenstates \citep{Avery:2013vwa}. The arguments of MP
do not hold in this situation.

The approach of \citep{Papadodimas:2013anh} may be considered a variant
on this theme. There one attempts to construct operators with the
properties of interior operators out of the complex late-time exterior
bulk operators. However this approach appears to fail consistency
checks. There is no explanation of why an interior observer would
experience quantum mechanics to a good approximation. The candidate
local interior operators do not commute with non-local exterior operators
at spacelike separations in the bulk. In the case of AdS/CFT, we believe
the time-evolution of these states by the CFT Hamiltonian leads them
to decohere into states that look local with respect to the usual
bulk exterior Hamiltonian at sufficiently late times. The local interactions
of the black hole with its exterior environment are of course what
is responsible for this decoherence. By choosing a sufficiently extreme
exterior environment, this decoherence time may be made quite short
\citep{Lowe:2014vfa}, of order a scrambling time. By the same token,
this local exterior bulk time evolution will then look highly non-local
from the viewpoint of these candidate interior operators. So while
the construction of such operators may appear to work at the level
of free field theory, it is not expected to survive the inclusion
of interactions.

\subsubsection{\label{sub:Unentangled-bulk-fields}Unentangled bulk fields}

An extension of the idea of section \ref{sub:No-interior-bulk} is
to assume in addition to the exterior bulk fields, one also has some
extra labels that represent the interior state of the bulk fields.
In view of the arguments of section \ref{sec:Euclidean-quantum-gravity}
this might seem reasonable, as one expects some set of extra labels
to account for the horizon entropy. However this runs counter to the
ideas of black hole complementarity, where one expects some kind of
different complementary description of the interior states, and that
only the exterior theory need have a conventional unitary description. 

In any case, this is the set of assumptions presented in MP, and if
one uses the resulting interior operators to build number operators
for infalling modes, one soon finds highly excited states in any typical
state in the microcanonical ensemble.

The two-sided black hole of figure \ref{fig:The-Penrose-diagram}
may be viewed as a particular example of this approach. The CFT factors
into two disjoint products $CFT_{L}\times CFT_{R}$. The construction
of local bulk interior operators in this framework has been described
in \citep{Hamilton:2006fh,Hamilton:2007wj,Lowe:2009mq}. The microcanonical
ensemble might be defined not just counting states of fixed energy
in $CFT_{R}$, taking the right region to represent the black hole
exterior, but also taking arbitrary energy in $CFT_{L}$ or any distribution
thereof. Because the propagation of states into the interior involve
potentially arbitrary states propagating from the left asymptotic
region, the future interior region will typically contain a firewall.

\subsubsection{\label{sub:Ancillae}Ancillae}

This approach may be viewed as an extension of the approach of section
\ref{sub:Entangled-bulk-fields} to define interior bulk operators
at arbitrary times after black hole formation. This procedure is described
in detail in \citep{Lowe:2014vfa}. The basic elements are the exact
state in the CFT, mapped to some exterior bulk state on a family of
timeslices centered on some infalling geodesic at horizon crossing.
By doing propagation back of order a scrambling time $\beta\log S$,
combined with the introduction of additional modes in an entangled
pure state (commonly called ancillae in the quantum computing literature),
which describe sub-Planck length modes in their vacuum state. Such
a construction is known as an isometry, because the norm of the states
is preserved, despite the dimension of the Hilbert space being enlarged.
One can propagate this state forward in time, back to the horizon
crossing time and beyond, to construct the quantum state and associated
local operators in a complementary theory relevant for an infalling
observer falling into the interior. 

In this picture, the modes with wavelengths shorter than the temperature
scale $\beta$ (at the horizon crossing time and beyond) will arise
directly from the vacuum ancillae modes. Thus no firewall is expected.

\section{Discussion and conclusions}

The approach of \ref{sub:No-interior-bulk} is the minimal interpretation
of the CFT data. As described in \ref{sub:Entangled-bulk-fields}
it fails for tiny fractions of the time, for a large AdS black hole.
But nevertheless is the simplest interpretation that works at typical
times. Let is consider the microcanonical ensemble of such black hole
states. Certainly there is not a firewall in the exterior by the arguments
of \citep{Lowe:2013zxa}. Local operators outside the horizon, in
a typical pure state will see deviations from the purely thermal global
vacuum result at order $e^{-S}$. By construction, one is prevented
from asking the question whether there is a firewall in the interior.

In the approach of \ref{sub:Unentangled-bulk-fields} one is able
to build interior bulk operators. However since the construction then
produces a firewall on the horizon, it is not self-consistent, since
the construction of the bulk operators presumed a smooth semiclassical
geometry around which to build operators describing small perturbations.
Perhaps one should then conclude that rather than proving the existence
of firewalls for typical states, that one's construction of interior
operators has failed.

Of the approaches considered here, the only one that appears to be
self-consistent is that of section \ref{sub:Ancillae} which produces
no firewall at typical times and for a typical state in the microcanonical
ensemble.

\paragraph*{Acknowledgements}

D.L. thanks Larus Thorlacius for discussions. This research was supported
in part by DOE grant DE-SC0010010 and an FQXi grant. This work was
also supported in part by the National Science Foundation under Grant
No. PHYS-1066293 and the hospitality of the Aspen Center for Physics.

\bibliographystyle{apsrev}
\bibliography{firewall2}

\begin{thebibliography}{14}
\expandafter\ifx\csname natexlab\endcsname\relax\def\natexlab#1{#1}\fi
\expandafter\ifx\csname bibnamefont\endcsname\relax
  \def\bibnamefont#1{#1}\fi
\expandafter\ifx\csname bibfnamefont\endcsname\relax
  \def\bibfnamefont#1{#1}\fi
\expandafter\ifx\csname citenamefont\endcsname\relax
  \def\citenamefont#1{#1}\fi
\expandafter\ifx\csname url\endcsname\relax
  \def\url#1{\texttt{#1}}\fi
\expandafter\ifx\csname urlprefix\endcsname\relax\def\urlprefix{URL }\fi
\providecommand{\bibinfo}[2]{#2}
\providecommand{\eprint}[2][]{\url{#2}}

\bibitem[{\citenamefont{Boulware}(1975)}]{Boulware:1974dm}
\bibinfo{author}{\bibfnamefont{D.~G.} \bibnamefont{Boulware}},
  \bibinfo{journal}{Phys.Rev.} \textbf{\bibinfo{volume}{D11}},
  \bibinfo{pages}{1404} (\bibinfo{year}{1975}).

\bibitem[{\citenamefont{Hartle and Hawking}(1976)}]{Hartle:1976tp}
\bibinfo{author}{\bibfnamefont{J.}~\bibnamefont{Hartle}} \bibnamefont{and}
  \bibinfo{author}{\bibfnamefont{S.}~\bibnamefont{Hawking}},
  \bibinfo{journal}{Phys.Rev.} \textbf{\bibinfo{volume}{D13}},
  \bibinfo{pages}{2188} (\bibinfo{year}{1976}).

\bibitem[{\citenamefont{Hawking and Page}(1983)}]{Hawking:1982dh}
\bibinfo{author}{\bibfnamefont{S.}~\bibnamefont{Hawking}} \bibnamefont{and}
  \bibinfo{author}{\bibfnamefont{D.~N.} \bibnamefont{Page}},
  \bibinfo{journal}{Commun.Math.Phys.} \textbf{\bibinfo{volume}{87}},
  \bibinfo{pages}{577} (\bibinfo{year}{1983}).

\bibitem[{\citenamefont{Marolf and Polchinski}(2013)}]{Marolf:2013dba}
\bibinfo{author}{\bibfnamefont{D.}~\bibnamefont{Marolf}} \bibnamefont{and}
  \bibinfo{author}{\bibfnamefont{J.}~\bibnamefont{Polchinski}},
  \bibinfo{journal}{Phys.Rev.Lett.} \textbf{\bibinfo{volume}{111}},
  \bibinfo{pages}{171301} (\bibinfo{year}{2013}), \eprint{1307.4706}.

\bibitem[{\citenamefont{Unruh}(1976)}]{Unruh:1976db}
\bibinfo{author}{\bibfnamefont{W.}~\bibnamefont{Unruh}},
  \bibinfo{journal}{Phys.Rev.} \textbf{\bibinfo{volume}{D14}},
  \bibinfo{pages}{870} (\bibinfo{year}{1976}).

\bibitem[{\citenamefont{Gubser et~al.}(1996)\citenamefont{Gubser, Klebanov, and
  Peet}}]{Gubser:1996de}
\bibinfo{author}{\bibfnamefont{S.}~\bibnamefont{Gubser}},
  \bibinfo{author}{\bibfnamefont{I.~R.} \bibnamefont{Klebanov}},
  \bibnamefont{and} \bibinfo{author}{\bibfnamefont{A.}~\bibnamefont{Peet}},
  \bibinfo{journal}{Phys.Rev.} \textbf{\bibinfo{volume}{D54}},
  \bibinfo{pages}{3915} (\bibinfo{year}{1996}), \eprint{hep-th/9602135}.

\bibitem[{\citenamefont{Lowe}(2009)}]{Lowe:2009mq}
\bibinfo{author}{\bibfnamefont{D.~A.} \bibnamefont{Lowe}},
  \bibinfo{journal}{Phys.Rev.} \textbf{\bibinfo{volume}{D79}},
  \bibinfo{pages}{106008} (\bibinfo{year}{2009}), \eprint{0903.1063}.

\bibitem[{\citenamefont{Hamilton
  et~al.}(2007{\natexlab{a}})\citenamefont{Hamilton, Kabat, Lifschytz, and
  Lowe}}]{Hamilton:2006fh}
\bibinfo{author}{\bibfnamefont{A.}~\bibnamefont{Hamilton}},
  \bibinfo{author}{\bibfnamefont{D.~N.} \bibnamefont{Kabat}},
  \bibinfo{author}{\bibfnamefont{G.}~\bibnamefont{Lifschytz}},
  \bibnamefont{and} \bibinfo{author}{\bibfnamefont{D.~A.} \bibnamefont{Lowe}},
  \bibinfo{journal}{Phys.Rev.} \textbf{\bibinfo{volume}{D75}},
  \bibinfo{pages}{106001} (\bibinfo{year}{2007}{\natexlab{a}}),
  \eprint{hep-th/0612053}.

\bibitem[{\citenamefont{Hamilton
  et~al.}(2007{\natexlab{b}})\citenamefont{Hamilton, Kabat, Lifschytz, and
  Lowe}}]{Hamilton:2007wj}
\bibinfo{author}{\bibfnamefont{A.}~\bibnamefont{Hamilton}},
  \bibinfo{author}{\bibfnamefont{D.~N.} \bibnamefont{Kabat}},
  \bibinfo{author}{\bibfnamefont{G.}~\bibnamefont{Lifschytz}},
  \bibnamefont{and} \bibinfo{author}{\bibfnamefont{D.~A.} \bibnamefont{Lowe}}
  (\bibinfo{year}{2007}{\natexlab{b}}), \eprint{0710.4334}.

\bibitem[{\citenamefont{Avery and Lowe}(2013)}]{Avery:2013vwa}
\bibinfo{author}{\bibfnamefont{S.~G.} \bibnamefont{Avery}} \bibnamefont{and}
  \bibinfo{author}{\bibfnamefont{D.~A.} \bibnamefont{Lowe}}
  (\bibinfo{year}{2013}), \eprint{1310.7999}.

\bibitem[{\citenamefont{Lowe and Thorlacius}(2014)}]{Lowe:2014vfa}
\bibinfo{author}{\bibfnamefont{D.~A.} \bibnamefont{Lowe}} \bibnamefont{and}
  \bibinfo{author}{\bibfnamefont{L.}~\bibnamefont{Thorlacius}},
  \bibinfo{journal}{Phys.Lett.} \textbf{\bibinfo{volume}{B737}},
  \bibinfo{pages}{320} (\bibinfo{year}{2014}), \eprint{1402.4545}.

\bibitem[{\citenamefont{Shenker and Stanford}(2014)}]{Shenker:2013pqa}
\bibinfo{author}{\bibfnamefont{S.~H.} \bibnamefont{Shenker}} \bibnamefont{and}
  \bibinfo{author}{\bibfnamefont{D.}~\bibnamefont{Stanford}},
  \bibinfo{journal}{JHEP} \textbf{\bibinfo{volume}{1403}}, \bibinfo{pages}{067}
  (\bibinfo{year}{2014}), \eprint{1306.0622}.

\bibitem[{\citenamefont{Papadodimas and Raju}(2014)}]{Papadodimas:2013anh}
\bibinfo{author}{\bibfnamefont{K.}~\bibnamefont{Papadodimas}} \bibnamefont{and}
  \bibinfo{author}{\bibfnamefont{S.}~\bibnamefont{Raju}},
  \bibinfo{journal}{Phys.Rev.Lett.} \textbf{\bibinfo{volume}{112}},
  \bibinfo{pages}{051301} (\bibinfo{year}{2014}), \eprint{1310.6334}.

\bibitem[{\citenamefont{Lowe and Thorlacius}(2013)}]{Lowe:2013zxa}
\bibinfo{author}{\bibfnamefont{D.~A.} \bibnamefont{Lowe}} \bibnamefont{and}
  \bibinfo{author}{\bibfnamefont{L.}~\bibnamefont{Thorlacius}},
  \bibinfo{journal}{Phys.Rev.} \textbf{\bibinfo{volume}{D88}},
  \bibinfo{pages}{044012} (\bibinfo{year}{2013}), \eprint{1305.7459}.

\end{thebibliography}

\end{document}